\documentclass[pre,twocolumn,showpacs,preprintnumbers,amsmath,amssymb]{revtex4}
\usepackage{mathptmx}
\usepackage{graphicx}
\usepackage{dcolumn}
\usepackage{bm}

\begin{document}

\newcommand{\beq}{\begin{equation}}
\newcommand{\eeq}{  \end{equation}}
\newcommand{\bea}{\begin{eqnarray}}
\newcommand{\eea}{  \end{eqnarray}}
\newcommand{\bit}{\begin{itemize}}
\newcommand{\eit}{  \end{itemize}}

\title{Universal response of quantum systems with chaotic dynamics}

\author{Diego A. Wisniacki$^{1}$}
\author{Natalia Ares$^{1}$}
\author{Eduardo G. Vergini $^{2,3}$}

\affiliation{$^1$Departamento de F\'isica,
FCEyN, UBA, Ciudad Universitaria, Buenos Aires C1428EGA, Argentina \\
$^2$Departamento de F\'isica,  Comisi{\'o}n Nacional de
Energ{\'i}a At{\'o}mica., Av. Libertador 8250,
Buenos Aires, Argentina \\
$3$ Departamento de F{\'i}sica, E.T.S.I. Agr\'onomos, Universidad Polit\'ecnica
de Madrid, 28040-Madrid, Spain}

\date{\today}

\begin{abstract}
The prediction of  the response of
a closed system to external perturbations
is one of the central problems in quantum mechanics, and in this respect, the local density of states (LDOS)
provides a deep description of such a response. The LDOS is the distribution of the overlaps
squared connecting the set of eigenfunctions with the perturbed one. 
Here, we show that in the case of closed systems with classically chaotic dynamics,
the LDOS is  a Breit-Wigner distribution under very general
perturbations of arbitrary high intensity. Consequently, we derive a semiclassical expression 
for the width of the LDOS which is shown to be very accurate for
paradigmatic systems of quantum chaos. 
This work demonstrates the universal response of quantum systems with classically chaotic dynamics.
\end{abstract}

\maketitle
The action of a perturbation on eigenfrequencies and eigenfunctions of a quantum system 
has been a subject of  paramount importance since the beginning of quantum theory.  Its 
understanding  is at the heart of fundamental problems of quantum mechanics like dissipation, 
phase transition or irreversibility. The usual perturbation theory is a good starting point to
describes successfully this effect when the perturbation is small. However, approximated 
theories  usually fail for strong perturbations and highly demanding computational  methods 
are needed to describe  characteristics of the perturbed system. 
\par
The local density of states (LDOS) or  {\it strenght function} is a widely studied magnitude 
to characterize the effect of perturbations on quantum systems and has been extensively 
computed for different systems and perturbations \cite{wigner,Flambaum, casati,Fyodorov,doron2}. The
LDOS is a distribution of the overlaps squared between
the unperturbed and perturbed eigenstates. Let us consider  a chaotic one parameter
dependent  Hamiltonian $H(x)$, and its quantum counterpart  $\hat{H}(x)$ with  eigenfrequencies 
$\omega_j(x)$ and eigenstates $\vert j (x)\rangle$.
Then, the LDOS of an eigenstate $\vert i (x_0)\rangle$ (that we call unperturbed) is given by,
\[
\rho_i(\omega,\delta x)= \sum_{j} \vert\langle j (x)\vert  i 
(x_0)\rangle\vert^{2} \delta (\omega -\omega_{i j}),
\]
with $\delta x=x-x_0$ and $\omega_{i j}=\omega_j(x)-\omega_i(x_0)$. Furthermore,
to avoid a dependence on the particular characteristics of the state $\vert i (x_0)\rangle$, an average 
over $n$ unperturbed states in a small frequency window is performed
\begin{equation}
\bar{\rho}(\omega,\delta x)= \frac{1}{n} \sum \rho_i(\omega,\delta x).
\label{ldos}
\end{equation}   
\par
The LDOS is related with important measures of irreversibility and sensitivity to perturbations 
in quantum systems as the survival
probability and  the Loschmidt echo (LE)  \cite{Jalabert-Pastawski, prosen-rev,jacquod-rev}. 
In fact, the LDOS is  the Fourier tranform of the survival probability \cite{wisniacki-cohen} and its
width  gives the decay rate of the LE for a small enough strength of the perturbation 
\cite{prosen-rev,jacquod-rev}. 
In this letter these relations are exploited to show that LDOS has  Lorentzian shape, usually called the
Breit-Wigner distribution, under very general
perturbations of arbitrary high intensity. Moreover, we derive a semiclassical expression for the
width of the LDOS, $\sigma_{sc}$, in chaotic systems.  
The derived expression only depends on the perturbation, while the properties of the system are taken 
into account through a uniform measure in phase space.
Although $\sigma_{sc}$ is derived for local perturbations we show
that it also works in the case of global perturbations as a consequence of the Lorentzian character 
of the LDOS, and the requirement of statistical independence between perturbed and unperturbed 
eigenfunction sets. Of course, such a requirement imposes restrictions on the admitted perturbations
and we discuss this point at the end of the letter.   
We test the ability of  $\sigma_{sc}$  to predict the width of the LDOS in  perturbed cat maps and the 
Bunimovich stadium billiard with boundary deformations, observing that it works 
very well in both systems, even for strong perturbations far away from the Fermi Golden Rule regime. 
These results demonstrate for the first time  the universal response of quantum 
chaotic systems to perturbations of classical nature. 
\par
The Fourier transform of Eq. (\ref{ldos}) is given by,
\begin{equation}
{\cal{F}}[\bar{\rho}](t,\delta x)=\frac{1}{n}  \sum e^{-{\rm i} \omega_i(x_0) t}  
\langle i (x_0)\vert e^{{\rm i} \hat{H}(x) t/\hbar}   
\vert i (x_0)\rangle,
\label{four}
\end{equation}  
where the sum runs over the amplitude fidelity of eigenstates (whose square modulus is the survival probability). 
Let us evaluate the previous sum  semiclassically.  Van\'{\i}cek has proposed an approximation 
of the amplitude fidelity, named the {\it dephasing
representation} \cite{Vanicek},  by assuming a classically small perturbation in such a way that 
the {\it shadowing} theorem \cite{hammel} is valid. Using such an approximation, Eq. (\ref{four}) leads to,
\begin{equation}
{\cal{F}}[\bar{\rho}](t,\delta x) \approx  \int dq dp W(q,p) \exp[-{\rm i} \Delta S_{t}(q,p,\delta x)/ \hbar] ,
\label{vani}
\end{equation}
where $\Delta S_{t}(q,p,\delta x)$ is the action difference  evaluated along the unperturbed orbit starting at 
$(q,p)$ that evolves a time $t$. Moreover, $W(q,p)= (1/n) \sum W_i(q,p) $, with $W_i(q,p)$ being 
the Wigner function of  $\vert i (x_0)\rangle$.
In chaotic systems, $W(q,p)$ reduces to a uniform distribution.  
\par
In the case of local perturbations, the right hand side of Eq. (\ref{vani}) has been evaluated on a Poincar\'e 
surface of section by Goussev {\it et al.} \cite{Richter 2008} resulting,
\begin{equation}
{\cal{F}}[\bar{\rho}](t,\delta x) \approx  e^{-\gamma |t|},
\label{gouss}
\end{equation}
with  
\begin{equation}
\gamma =  \eta \left(1- \Re  \left\langle e^{-{\rm i} \Delta S(q,p, \delta x)/\hbar} \right\rangle \right).
\label{alpha}
\end{equation}
\begin{figure}[t]
\begin{center}
\includegraphics[width=8.cm] {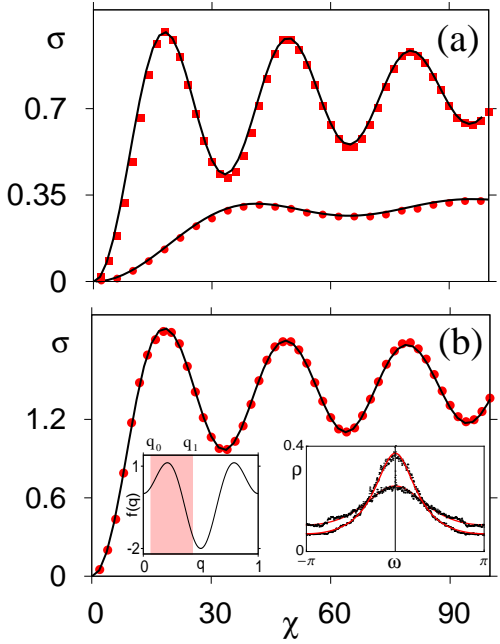}
\end{center}
\caption{(Color online) Width $\sigma$ of the LDOS as a function of the scaled perturbation strength
$\chi = N \delta k $, for a local perturbation. Solid symbols correspond to the quantum 
case and solid lines to the semiclassical calculation $\sigma^{(p)}_{sc}$. We use  $N=800$ and $q_0=0.01$. 
In panel (a) we use
a width $\beta=0.2$ (circles) and 
$\beta=0.4$ (squares). In panel (b) $\beta=0.7$.
Left inset:  Schematic figure showing  the used  local perturbation. The scaled 
shear [$f(q)\equiv 2 \pi \epsilon(q,k)/k$, (with $k$ the strength of the perturbation)] is plotted as a function of $q$. 
The limits of the perturbed region are indicated with $q_0$ and $q_1$, being $\beta=q_1-q_0$ its width.
Right inset: The LDOS $\rho$ for $\beta=0.7$, with $\chi=78.4$ and $89.6$. In solid (red) lines we plot  periodized 
Lorentzian functions with the corresponding widths. }
\label{fig1}
\end{figure}
The average  is evaluated on the region of surface of section where the local perturbation acts,
and $ \Delta S(q,p, \delta x)$ is the action difference after one step on the surface of section.
For numerical comparison we consider a rectangular region,
\begin{equation}
\left\langle e^{-{\rm i} \Delta S(q,p, \delta x)/\hbar} \right\rangle = \frac{1}{\alpha} 
\int_{p_1}^{p_2}  \int_{q_1}^{q_2}  e^{-{\rm i} \Delta S(q,p, \delta x)/\hbar} dq dp,
\label{prom}
\end{equation} 
\noindent
where $p_1$, $p_2$, $q_1$ and $q_2$ are the limits of the perturbed region with 
area $\alpha=(p_2-p_1) (q_2-q_1)$. Moreover, $\eta$, named the classical decay rate in 
Ref. \cite{Richter 2008},  is the 
probability  to reach the perturbed region per unit time,
\begin{equation}
\eta =  \frac{\alpha}{{\tau \cal A}},
\label{eta}
\end{equation}
where $\cal A$ is the area of the Poincar\'e  surface of section  and $\tau$ is the mean mapping time.  
\par
The inverse Fourier tranform of Eq. (\ref{gouss})  
is a  Lorentzian function, the so-called Breit-Wigner distribution,
\begin{equation}
\bar{\rho}(\omega,\delta x)\approx L(\gamma,\omega)=
\frac{\gamma}{\pi (\omega^2+\gamma^2)},
\label{xxxx}
\end{equation}
and we define its width as
half the distance around the mean value that contains the $70 \%$ of the probability (actually, this value is
only relevant for the numerical computation). Then, 
the semiclassical approximation of the width results,
\begin{equation}
\sigma_{sc} =  \tan \left( 0.7\frac{\pi}{2} \right) \gamma \approx 1.963 \gamma.
\label{prop2}
\end{equation}   
\par
We stress out that  the semiclassical approximation was derived in the limit of $\alpha \rightarrow 0$. 
However, our final expression (Eq. (\ref{prop2})) can be 
extended to arbitrary values of $\alpha$ as a result of the Lorentzian character of the LDOS and
the property of short range correlation of chaotic eigenfunctions.
To clarify this point, let us consider the following
 basis sets:  $\{\vert i ^{(0)}\rangle \}$ be the set of unperturbed eigenfunctions,  $\{\vert i ^{(1)}\rangle \}$  be 
 the set resulting after applying a local perturbation over an infinitesimal
region $\delta \alpha_1$, and   $\{\vert i ^{(2)}\rangle \}$  be the set resulting after applying the 
perturbation $\delta \alpha_2$
to the previous system (the one with the perturbation  $\delta \alpha_1$). For $\delta \alpha_1$ and
$\delta \alpha_2$ being perturbations over disjointed regions of phase space,  their corresponding LDOS should be   
statistical independent because chaotic eigenfunctions have correlations  of short range in phase space. 
Therefore by assuming independence, the LDOS connecting the first and third basis sets 
is simply derived by  convoluting the previous ones,
\[
\int  L(\delta \gamma_1, \omega_1)  L(\delta \gamma_2, \omega-\omega_1)  {\rm d} \omega_1 = 
L(\delta \gamma_1+\delta \gamma_2,\omega).
\]
The new LDOS is also a Lorentzian function, 
with $\delta \gamma_1$ ($\delta \gamma_2$) being the parameter corresponding to the first (second) perturbation.
Then, by using Eqs. (\ref{alpha}), (\ref{prom}) and (\ref{eta}) it is easy to see that $\delta \gamma_1 +\delta \gamma_2$ 
is just the parameter 
resulting from the perturbation $\delta \alpha_1+\delta \alpha_2$. Following this procedure, we can
now add a perturbation  $\delta \alpha_3$ and so on up to fill a finite region $\alpha$. 
We would like to stress that even though the property of statistical independence is very reasonable, there
are exceptional perturbations where such a property is not satisfied; see the end of the letter for a discussion
of this point.
\par
\begin{figure}
\setlength{\unitlength}{1cm}
\begin{center}
\includegraphics[width=7. cm] {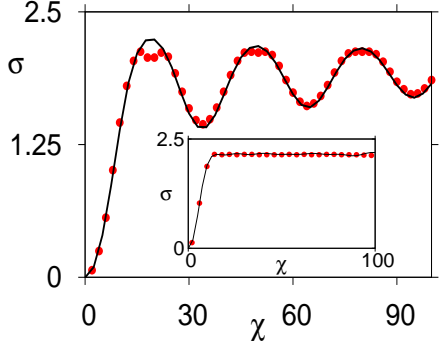}
\end{center}
\caption{ Idem Fig.~\ref{fig1} for global perturbations. 
The perturbation in the main plot is the same as in Fig.~\ref{fig1}. 
For the inset, the perturbation is a 
shear in momentum and position (see text for details). }
\label{fig2}
\end{figure}
Let us show the power of  the semiclassical approximation to describe  the LDOS
in quantum maps, where the mapping time is fixed to $\tau =1$, the phase space  is a torus of area
${\cal{A} }= 1$, and the Hilbert space has finite dimension $N$ (with $2 \pi \hbar =1/N$). We 
consider the cat map, a canonical example in classical and quantum chaos studies, 
perturbed with a non-linear shear in momentum,
\begin{equation}
\begin{array}{lcl}
       q' & = & 2q + p  \nonumber \\
       p' & = & 3q + 2p + \epsilon(q,k) \nonumber \\
 \end{array}
\  ({\rm mod}\ 1),
 \end{equation}
where $\epsilon (q,k) =(k/2\pi) [\cos(2\pi q)-\cos(4\pi q)]$, with $k$ being the strength of the perturbation. 
The action difference  for one iteration of the map is given by 
$\Delta S(q,\delta k)=(\delta k/4\pi^{2})[\sin(2\pi q)-\sin(4\pi q)/2]$ \cite{Ozorio 1994}.
\par
For local perturbations \cite{Ares}, the shear $\epsilon (q,k)$ is only applied to a $q$ window from $q_0$ to $q_1$, 
with width $\beta=q_1-q_0$ [see left inset in Fig.~\ref{fig1}]; so, 
$\alpha=\beta$. Furthermore, we take into account the fact that the spectrum of the cat map is periodic 
because of a compact phase space. 
This periodicity changes the form of the LDOS because the probability that leaves from one border returns to the other. 
By assuming no correlation between  the existing and returning probabilities, the LDOS transforms into a periodized
Lorentzian function 
\[
L^{(p)}(\gamma,\omega)=\sum_{j=-\infty}^{\infty} L(\gamma,\omega-2 \pi j/\tau),
\]
with $\omega \tau$ being the variable that specify the spectrum of eigenphases. 
This distribution provides a new relation between the width of the distribution and $\gamma$, whose first 
correction with respect to
Eq. (\ref{prop2}) is  $ \sigma_{sc}^{(p)} \simeq \sigma_{sc} [1-(\sigma_{sc}/\pi)^2)]$. On the other hand, a
numerical computation reveals a linear term which should be related to correlations between the existing and
returning probabilities.  We obtain the following estimate,
\begin{equation}
\sigma^{(p)}_{sc} \approx \sigma_{sc} [1+0.24 \sigma_{sc}-(\sigma_{sc}/\pi)^2],
\label{periodsigma}
\end{equation}
by fitting  the linear term to the numerical data.
\par
Fig.\ref{fig1} shows the width, $\sigma$, of the LDOS 
and its semiclassical approximation, $\sigma^{(p)}_{sc}$,
for three different windows in positions where the
 perturbation is applied. We plot the width as a function of the scaled perturbation strength,
$\chi =N \delta k $, in such a way that figures are insensitive to $N$. 
As it can be seen  the  semiclassical approximation works very well.
Moreover, the right inset of Fig.\ref{fig1} (b) shows that the LDOS is a periodized 
Lorentzian function even for strong perturbations  far  away from the Fermi 
Golden Rule regime, which is identified with the quadratic behavior close to the origin.
\par
Fig. \ref{fig2} compares the width of the LDOS with $\sigma^{(p)}_{sc}$ in the case of
global perturbations ($\alpha=1$). The perturbation in the main panel is the one 
used previously while in the
inset, we use the same shear in momentum plus the following shear
in position: $\bar{\epsilon}(p,k)=-(k/2\pi)[\sin(6\pi p)/3+\cos(4\pi p)/2]$. In the latter
the action difference is, 
$\Delta S(q,p,\delta k)=(\delta k/4\pi^{2})[\sin(2\pi q)-\sin(4\pi q)/2+
\cos(6\pi p)-\sin(4\pi p)/2] $. 
Differences near the peaks at
$\chi \approx 20$ and $\chi \approx 50$ are related to the poor accuracy of Eq. (\ref{periodsigma}) close
to the saturation, which is given by the width of the 
uniform distribution, $0.7 \pi \approx 2.20$.
\begin{figure}
\setlength{\unitlength}{1cm}
\label{fig-stadium}
\begin{center}
\includegraphics[width=6.cm] {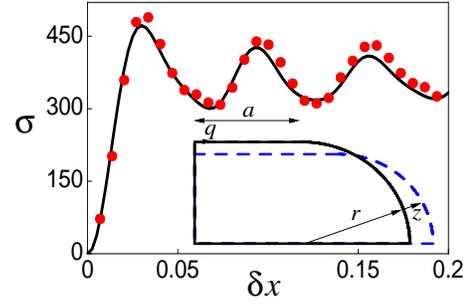}
\caption{$\sigma$ as a function of the perturbation strenght
$\delta x$ for the stadium billiard in the region of wave number around $200$ (solid circles). The semiclassical 
approximation $\sigma_{sc}$ using Eq. (\ref{prop2}) is plotted with solid line.  Inset: 
two Bunimovich stadium billiards with different shape parameter; the perturbed system is indicated
in dashed line.}
\label{fig-stadium}
\end{center}
\end{figure}
\par
To further demonstrate the power of the proposed semiclassical approximation in a realistic system, 
we consider the desymmetrized Bunimovich stadium billiard with radius $r$ and straight line of 
lenght $a$. 
This  system is fully 
chaotic \cite{cit-bunimovich} and has great theoretical and experimental relevance.
The billiard is perturbed by 
the boundary deformation displayed in the inset of Fig \ref{fig-stadium}. The area of the 
billiard is fixed to the value $1+\pi/4$, so the boundary only depends on
the shape parameter $x=a/r$. 
The boundary deformations are parametrized by
$
{\bf r} (q,\delta x )= {\bf r_{0}} (q)+z(q,\delta x )\;{\bf n},
$
where $q$ is a coordinate along the unperturbed boundary $\cal C$, 
${\bf r_{0}} (q)$ defines $\cal C$ and
${\bf n}$ is the outward normal unit vector to ${\cal C}$ at ${\bf r}_{0}(q)$ (an explicit expression for
$z(q,\delta x )$ is provided in Ref. \cite{wisniacki-vergini}).
We consider the usual Birkhoff coordinates
to describe the classical dynamics of the particle; that is, the variables $q$ and $p=\sin(\theta)$,
with $\theta$ the impinging angle with ${\bf n}$. To compute $\sigma_{sc}$, the action
difference between the unperturbed and perturbed orbit results in \cite{Richter 2008},
\[
\Delta S(q, \theta, \delta x)=\left | {\bf p} \right | \Delta L=2 \left | {\bf p} \right | z(q,\delta x) \cos( \theta),
\]
where $\Delta L$ is the lenght difference between the unperturbed and perturbed
orbits, and $ {\bf p}$ is the momentum of the particle. The mean time between bounces 
with the boundary is given by $\tau=m \pi A/(\left | {\bf p} \right | {\cal P})$ \cite{chernov}, 
with $m$ the mass 
of the particle, $A$ the area of the billiard and ${\cal P}$ its perimeter. Then, the decay rate results in
$
\eta=\left | {\bf p} \right | \beta/(m \pi A),
$
with $\beta$ the width of the perturbed region\cite{Richter 2008}; for the selected perturbation,
$\beta={\cal P}$.
Fig. \ref{fig-stadium} compares the numerical and semiclassical calculations;  $\hbar=1$ and $m=1/2$
are used.
The width computed with the exact 
eigenstates is plotted with full circles and the semiclassical aproximation, $\sigma_{sc}$, is plotted in full line.
The calculations  displayed in Fig. \ref{fig-stadium} were performed
around  the wave number $200$, where the eigenstates of the 
billiard were computed using the scaling method \cite{Vergini}. 
The agreement between the quantum and the semiclassical calculation is excellent. 
We notice that while the full quantum computation of 
$\sigma$ in Fig. \ref{fig-stadium} is very  time consuming 
[$t \approx 7 \times 10^7 seg$ in an CPU Intel Core 2 6400], 
the semiclassical calculation is a simple two variable integral. 
\par
One final point is  to discuss the character of the perturbation in order to
satisfy the used property of statistical independence  between the perturbed  and
unperturbed eigenfunction sets. Let us consider as 
an extreme example a chaotic Hamiltonian of the form kinetic plus 
potential energy, and where the perturbed Hamiltonian is obtained by a
displacement of the potential energy. In this situation, the two systems 
have the same spectrum and the corresponding eigenfunctions are
connected by the displacement. So, the two sets are strongly
correlated and the LDOS  does not satisfy the Breit-Wigner distribution; 
see for instance Ref. \cite{barnett} where this type of perturbations is analyzed. 
From the classical point of view, we notice that the dynamics and in particular, 
the structure of periodic orbits are not affected at all by the perturbation. 
In this context, the question that immediately arises is how the perturbation has
to modify the dynamics in order to guarantee the required independence. We develop the answer 
within the short periodic orbit approach \cite{short},  where the eigenfunctions of a 
chaotic system are
described in a scar function basis set \cite{carlo};  these wave functions are supported by 
the shortest periodic orbits of the system, with periods up to the Ehrenfest 
time. On the other hand, each matrix
element in this basis includes a phase depending on the difference of actions between 
the corresponding periodic orbits \cite{david}.
So, to get statistical independence  between the 
perturbed  and unperturbed eigenfunction sets, the perturbation has to
modify  the action of the used periodic orbits by at least $\hbar$, in a more or less random way.
Specifically, the set of numbers $[\Delta S_{\mu}/(2 \pi \hbar)]$ (mod $1$), where $\mu$ labels
periodic orbits with period shorter than the Ehrenfest  time, has to be distributed uniformly in the
interval $[0,1)$; $\Delta S_{\mu}$ is the action of the unperturbed periodic orbit, $\mu$, minus 
the corresponding action in the perturbed case. 
\par
In conclusion, our results demonstrate that quantum systems with classically chaotic dynamics react in a
universal way as a consequence of perturbations of classical nature.
Specifically, the LDOS is a Breit-Wigner distribution, even for strong perturbations.
Moreover,  we  derive a semiclassical expression for its width 
that  is accurate for paradigmatic systems of quantum chaos as 
the cat maps and the stadium billiard.
As a final remark, we would like to notice that our semiclassical result reproduces in part the old one 
obtained by Wigner \cite{wigner} within the random matrix theory \cite{note}. This fact implies that
the connection between chaotic systems and random matrix theory, uncovered by the cellebrated  
Bohigas-Giannoni-Schmit conjecture \cite{bohigas}, is stronger than believed.

\par
We acknowledge the support from CONICET (PIP-112-200801-01132) , UBACyT (X237), ANPCyT and 
MTM2009-14621. We would like to thank  Doron Cohen and Marcel Novaes for
useful discussions. 


\begin{thebibliography}{99}

\bibitem{wigner} E. P. Wigner. Ann. Math. {\bf 62}, 584 (1955).

\bibitem{Flambaum} V. V. Flambaum, A. A. Gribakina, G. F. Gribakin, and M. G. Kozlov.
Phys. Rev. A. {\bf 50}, 267 (1994).

\bibitem{Fyodorov} Y Fyodorov, O Chubykalo, F Izrailev and G Casati.
Phys. Rev. Lett. {\bf 70}, 1603 (1996).

\bibitem{casati} G. Casati , B.V. Chirikov, I. Guarneri, 
and M. Izrailev. 
 Phys. Lett. A {\bf 223}, 430 (1996).

\bibitem{doron2}
D. Cohen and E.J. Heller. Phys. Rev. Lett. {\bf 84}, 2841 (2000).

\bibitem{Jalabert-Pastawski} 
R.A. Jalabert and H.M. Pastawski.  Phys. Rev. Lett. {\bf 86}, 2490 (2001).

\bibitem{prosen-rev}
T. Gorin, T Prosen, TH Seligman and M. \u{Z}nidari\u{c}.  Phys. Rep. {\bf 435}, 33 (2006).

\bibitem{jacquod-rev}
Ph. Jacquod and  C. Petitjean. . Adv. in Phys. {\bf 58},  67 (2009).


\bibitem{wisniacki-cohen}
D.A. Wisniacki and D. Cohen. Phys. Rev. E {\bf 66}, 046209 (2002).


\bibitem{hammel} S. M. Hammel, J. A. Yorke and C. Grebogi. J. Complex. {\bf 3}, 136 (1987).

\bibitem{Vanicek} 
 J.Van\'{\i}cek.  Phys. Rev. E {\bf 70}, 055201 (R) (2004); {\it ibid}  {\bf 73}, 046204  (2006).

\bibitem{Richter 2008}
A.Goussev, D. Waltner, K. Richter and R. A. Jalabert. 
  New Journal of Physics {\bf 10}, 093010 (2008).

\bibitem{Ozorio 1994}
M.Basilio De Matos, A. M. Ozorio De Almeida.  Ann. Phys. {\bf 237}, 46-65 (1995).

\bibitem{Ares} N. Ares and D. A. Wisniacki.  Phys. Rev. E {\bf 80}, 046216 (2009).

\bibitem{cit-bunimovich}  L. A. Bunimovich.  Funct. Anal. Appl. {\bf 8} 
254 (1974).



\bibitem{wisniacki-vergini} D. A. Wisniacki and E. Vergini. Phys. Rev. E {\bf 59}, 6579 (1999).

\bibitem{chernov} N. Chernov.  J. Stat. Phys. {\bf 88}, 1 (1997).


\bibitem{Vergini} E. Vergini and M. Saraceno.  Phys. Rev. E {\bf 52}, 2204 (1995).

\bibitem{barnett} A. H. Barnett, D. Cohen, and E. J. Heller. Phys. Rev. Lett. {\bf 85}, 1412 (2000).

\bibitem{short} E. G. Vergini. J. Phys. A: Math. and Gen. {\bf 33}, 4709 (2000); E. G. Vergini, D. Schneider
and  A. M. F. Rivas. J. Phys. A: Math. Theor. {\bf 41}, 405102 (2008).

\bibitem{carlo}  E. G. Vergini and G. G. Carlo. J. Phys. A: Math. and Gen. {\bf 34}, 4525 (2001).

\bibitem{david} E. G. Vergini and D. Schneider. J.Phys. A: Math. and Gen. {\bf 38}, 587 (2005).

\bibitem{note} In the limit of infinite perturbations, Wigner find a transition from the Lorentzian shape to a 
semicircular law. We speculate that our development is unable to detect such a transition because the used
perturbations, even strong at quantum level, do not affect considerably the classical structure of the system.

\bibitem{bohigas} O. Bohigas, M. J. Giannoni and C. Schmit.  Phys. Rev. Lett. {\bf 52}, 1 (1984).
\end{thebibliography}
\end{document}